\input harvmac
\input epsf

\def\half{{1\over 2}}

% \def\lt{\tilde{\lambda}}

%%%%%%%%%%%%%%%%%%%%%%%%%%%%%%%%%%%%%%%%%%%%%%%%%%%%%%%%%%%%%%%%%%%%%

\Title{}{\vbox{\centerline{CMB Power Spectrum from Noncommutative Spacetime}}}

\centerline{Qing Guo Huang$^{1}$ and Miao Li$^{1,2}$ }
\centerline{$^{1}$\it Institute of Theoretical Physics}
\centerline{\it Academia Sinica, P.O. Box 2735}
\centerline{\it Beijing 100080} 
\medskip
\centerline{$^{2}$\it Department of Physics}
\centerline{\it National Taiwan University}
\centerline{\it Taipei 10764, Taiwan}
\centerline{\tt huangqg@itp.ac.cn}
\centerline{\tt mli@itp.ac.cn}

\bigskip

Very recent CMB data of WMAP offers an opportunity to test inflation
models, in particular, the running of spectral index is quite new and
can be used to rule out some models. We show that an noncommutative spacetime
inflation model gives a good explanation of these new results. In fitting
the data, we also obtain a relationship between the noncommutative 
parameter (string scale) and the ending time of inflation. 

\Date{April, 2003}

\nref\ben{C. L. Bennett et al., astro-ph/0302207.}
\nref\sper{D. N. Spergel et al., astro-ph/0302209.}
\nref\pei{H. V. P. Peiris et al., astro-ph/0302225.}
\nref\sind{B. Feng, M. Li, R.-J. Zhang and X. Zhang, astro-ph/0302479;
J. E. Lidsey and R. Tavakol, astro-ph/0304113;
M. Kawasaki, M. Yamaguchi and J. Yokoyama, hep-ph/0304161;
L. Pogosian, S.-H. H. Tye, I. Wasserman and M. Wyman, hep-th/0304188.}
\nref\bh{R. Brandenberger and P.-M. Ho, hep-th/0203119.}
\nref\ss{C-S. Chu, B. R. Greene and G. Shiu, hep-th/0011241;
F. Lizzi, G. Mangano, G. Miele and M. Peloso, hep-th/0203099.}
\nref\all{A. R. Liddle and D. H. Lyth, Cosmological Inflation and Large-
Scale Structure (Cambridge University Press, 2000).}
\nref\ty{T. Yoneya, in ``Wandering in Fields," eds. K. Kawarabayashi, A. Ukawa
(World Scientific, 1987); M. Li and T. Yoneya, Phys. Rev. Lett. 78,
(1997) 1219, hep-th/9611072; T. Yoneya, Prog. Theor. Phys. 103 (2000)
1081, hep-th/0004074.}

Recent results from the Wilkinson Microwave Anisotropy Probe \refs{\ben-\pei}
give further determination of cosmological parameters and constrain properties
of inflationary models tightly. Some analysis was done in \pei, and in particular,
the new data on the running of the spectral index of the power spectrum of CMB
provide good constraints on inflationary models, and some models were analyzed
in \sind. In general, models satisfying the slow roll conditions naturally
predict too small running of the spectral index. In this note we analyze
the model of \bh\ in which spacetime noncommutativity is taken into account,
and find that it is rather easy to accommodate WMAP data in this model.
It is quite remarkable that the two parameters in this model can be used
to fit four data of the spectral index. We determine these two parameters,
and the string scale is in turn determined once the termination time
of inflation is known. For a typical termination time of inflation $t_*=
10^{-32}\hbox{s}$ \all, the string scale is $l_s=10^{-25}$cm, or $M_s=10^{11}$
Gev.

Spacetime in general is noncommutative in string theory, this manifests in
a general relation $\Delta X\Delta T\ge l_s^2$ \ty. However it is hard to formulate
string theory in a fashion to incorporate this relation directly. On a general
ground, if inflation reflects physics at a scale close to string scale or
a related scale, one expects that spacetime uncertainty must have effects
in the CMB power spectrum, observable or to be observable (implications
of space-space noncommutativity for inflation were considered in \ss). 
Lacking a manifest
formulation of noncommutativity, we shall be content with a scheme proposed
in \bh, in which spacetime noncommutativity is put in by hand for a scalar
field (inflaton). Skipping the detailed discussions leading to their action,
we copy the free scalar action here
\eqn\ncscal{S=V\int d\eta d^3k z_k^2(\eta)(\phi^\prime_k\phi_{-k}^\prime
-k^2\phi_k
\phi_{-k}),}
where
\eqn\scaf{\eqalign{z_k^2&=a(\eta)^2y_k^2(\eta), \quad 
y_k^2=(\beta_k^+\beta_k^-)^{\half},\cr
{d\eta\over d\tau}&=\left({\beta_k^-\over \beta_k^+}\right)^\half,\quad
\beta_k^\pm =\half (a^{\pm 2}(\tau+\l_s^2k)+a^{\pm 2}(\tau-l_s^2k),}}
where $l_s$ is the string length scale, $a(\tau)$ is the scale factor in
the metric, and time $\tau$ is defined such that
\eqn\metr{ds^2=-a^{-2}(\tau)d\tau^2+a^2(\tau)dx^2.}
Unlike in the commuting spacetime, the all important scale factor $z_k$
in our case depends on the comoving wave number $k$. 
 
We shall consider the power-law inflation, and this can be generated by a 
scalar potential assuming an exponential form. Let $t$ be the cosmological 
time, so $a(t)=a_0t^n$, $n>1$. In terms of $\tau={a_0\over n+1}t^{n+1}$, 
$a=\alpha_0\tau^{{n\over n+1}}$ with
\eqn\repa{\alpha_0=a_0^{{1\over n+1}}(n+1)^{{n\over n+1}}.}
To simplify relations that will follow, we define a scale $l$ so that
$a(\tau)=(\tau/l)^{{n\over n+1}}$ and $a(t)=({{t\over (n+1)l}})^n$.
We shall estimate the relationship between the scale $l$ and the termination
time $t_*$ later.

The scalar power spectrum can be computed in the usual way, and is given by
\eqn\powers{P(k)={k^2\over 4\pi^2z_k^2(\eta_k)},}
and $\eta_k$ is determined by the condition
\eqn\tkd{k^2={z_k''\over z_k}.}
Two extremal limits were considered in \bh, namely the UV limit and the IR
limit. However, the observed results must lie in between, so we need to calculate
the power spectrum again. The interested region is actually UV in terms of
the time when the perturbation is created, but this was not discussed in \bh.
What we need to compute is 
\eqn\sind{n_s=1+{d\ln P(k)\over d\ln k},\quad {dn_s\over d\ln k}.}

Define
\eqn\kcr{k_c=\left({(n+1)^2\over n(2n-1)}\right)^{-{n+1\over 4}}({l_s\over l})^{n-1}
l^{-1}.}
Naively, $k_c$ is determined by two microscopic scales $l$ and $l_s$, one would
expect that $k_c$ is also at the microscopic scale. This is wrong, due to the large
exponent $n$. To achieve sufficient number of e-folds, $n$ ought to be large
enough, and indeed we will determine it be around 13. Thus, if $l_s$ is smaller
than $l$ by a few order, the characteristic wave length $1/k_c$ is macroscopic.
The interesting range of $k$ is between $(0.002, 0.05)\hbox{Mpc}^{-1}$, and we
assume these length scales be smaller than $1/k_c$, so $k_c/k\ll 1$. In solving
the crossing horizon condition \tkd, we will expand everything in $k_c/k$.

Without the string spacetime uncertainty, or let $l_s=0$, the crossing horizon
time is
\eqn\nscr{\tau_k^0=l_s^2k\left({k\over k_c}\right)^{{2\over n-1}},}
the dependence on $l_s$ is fictitious, since a factor in $k_c$ cancels $l_s^2$.
Since string uncertainty introduces a uncertainty in time $\sigma=l_s^2k$,
this ought to be smaller than the time when the perturbation is created,
namely, $\sigma/\tau_k^0$ must be small. This ratio is nothing but
\eqn\rat{{\sigma\over \tau_k^0}=\left({k_c\over k}\right)^{{2\over n-1}}\ll 1.}
For a large enough $n$, $k_c$ is small enough so that the above assumption
will be valid.

In solving \tkd, not only we need to modify $z_k$ according to \scaf, but also 
we need to modify the definition of the conformal time according to \scaf.
For instance
\eqn\deri{{d\over d\eta}=\alpha_0^2\tau^{{2n\over n+1}}(1-{n\over n+1}{\sigma^2
\over\tau^2}){d\over d\tau},}
where we have truncated to the first nontrivial order in $\sigma/\tau$. After a lengthy
but straightforward calculation, we find
\eqn\modsca{{z_k''\over z_k}={n(2n-1)\over (n+1)^2}\alpha_0^4\tau^{{2(n-1)\over n+1}}
\left(1-{2n(n+1)(2n-5)\over (2n-1)(n+1)^2}{\sigma^2\over\tau^2}\right).}
Solving \tkd\ using \modsca, we obtain
\eqn\crti{\tau_k=l_s^2k({k\over k_c})^{{2\over n-1}}\left(1+{n(2n-5)\over (n-1)(2n-1)}
({k_c\over k})^{{4\over n-1}}\right).}
The power spectrum is readily calculated to be
\eqn\pwsp{P(k)\sim k^{-{2\over n-1}}\left(1-{4n^2(n-2)(2n+1)\over (n+1)^2(n-1)
(2n-1)}({k_c\over k})^{{4\over n-1}}\right).}
Denote the coefficient in the front of $(k_c/k)^{{4\over n-1}}$ in the above formula
by $x$ (the positive one), we compute the spectral index and the running of the index
\eqn\inrin{\eqalign{n_s-1&=-{2\over n-1}+{4x\over n-1}({k_c\over k})^{{4\over n-1}},\cr
{dn_s\over d\ln k}&=-{16x\over (n-1)^2}({k_c\over k})^{{4\over n-1}}.}}
Since $x$ is positive, $n_s$ is larger for smaller $k$, the expected behavior.
Now, use the WMAP data
\eqn\wmap{\eqalign{n_s&=0.93\pm 0.03, \quad  {dn_s\over d\ln k}=-0.031^{+0.016}
_{-0.017}
\quad \hbox{at}\quad k=0.05 \hbox{Mpc}^{-1},\cr
n_s&=1.10^{+ 0.07}_{-0.06}, \quad  {dn_s\over d\ln k}=-0.042^{+0.021}_{-0.020}
\quad \hbox{at}\quad k=0.002 \hbox{Mpc}^{-1}}}
to see whether the main formulas in \inrin\ can fit these. There are only two
parameters in \inrin, namely $n$ and $k_c$, but there are four independent
results from the WMAP. We will adopt the following strategy: we first use the
two results at $k=0.05 \hbox{Mpc}^{-1}$ to determine the two parameters, and
make a prediction about $n_s$ and its running at $k=0.002 \hbox{Mpc}^{-1}$.
Later, we use the results about $n_s$ at two different scales to make a prediction
about the running of $n_s$ at the two scales.

Using the results at $k=0.05 \hbox{Mpc}^{-1}$, we find
\eqn\ffit{n=13.171^{+6.198}_{-2.772},\quad k_c=1.65^{+6.84}_{-1.64}\times 10^{-5}
\hbox{Mpc}^{-1}.}
Indeed, $k_c/k$ is a small number, and $(k_c/k)^{{4\over n-1}}$ is also small
enough so the approximation used in deriving \inrin\ is valid.
These parameters together with \inrin\ predict
\eqn\irpre{n_s=1.11^{+ 0.14}_{-0.13}, \quad  {dn_s\over d\ln k}=-0.089^{+0.059}
_{-0.100}
\quad \hbox{at}\quad k=0.002 \hbox{Mpc}^{-1}.}
The central value of the predicted $n_s$ is quite good compared to that in
\wmap, the central value of the predicted running is larger than that in \wmap,
but within the error bar, we notice that the tendency of the predicted
running is good, it is larger at a larger scale. For $k=0.002\hbox{Mpc}^{-1}$,
it is possible that higher orders in $k_c/k$ are not negligible, and may improve
our result. We also notice that in \pei\ the central value of the running is
estimated to be $-0.077$, much closer to our prediction. 

Next, using the values of $n_s$ at two different scales in \wmap, we determine
\eqn\sfit{n=13.336^{+5.204}_{-2.412},\quad k_c=1.43^{+3.81}_{-1.42}\times 10^{-5}
\hbox{Mpc}^{-1},}
and the running of the spectral index
\eqn\irpree{{dn_s\over d\ln k}(k=0.05\hbox{Mpc}^{-1})=0.030^{+ 0.013}_{-0.011}, 
\quad  {dn_s\over d\ln k}(k=0.002\hbox{Mpc}^{-1})=-0.085^{+0.050}
_{-0.065}.}

We show the spectral index and the running as functions of $\ln k$ in fig.1
and fig.2.

We have used the WMAP data to determine two parameters $n$ and $k_c$ in 
the noncommutative inflation model. Next, we want to determine the relation
between the ending time of inflation and the string scale. To do so, note
that the comoving wave numbers are measured at the present time, so the normalization
of the scale factor $a(t)$ at the present $t_0$ is 1. Since the termination
of inflation, our universe underwent a radiation dominated epoch and matter
dominated epoch (for simplicity we ignore the observed dark energy), so
the scale factor was enhanced twice:
\eqn\enh{a(t_*)({10^4\times 365\times 24\times 60^2\over t_*})^{\half}
({13.7\times 10^9\over 10^4})^{{2\over 3}}=1,}
where $t_*$ is the termination time of inflation and its unit is second.
Using $a(t_*)=a_0t_*^n$, we find the parameter $l$ 
\eqn\pasca{l={t_*\over n+1}({4.8\times 10^9\over t_*})^{{1\over 2n}}.}
$l$ is larger than $t_*$. Next, using the definition of $k_c$ in \kcr, we get
\eqn\strs{l_s=\left[k_cl^n\left({(n+1)^2\over n(2n-1)}\right)^{{n+1\over 4}}
\right]^{{1\over n-1}}.}
In the end, $l$ and $l_s$ are expressed in terms of $t_*$
\eqn\scarel{l=dt_*^{{2n-1\over 2n}},\quad l_s=ct_*^{{2n-1\over 2n-2}},}
where in defining $d$ and $c$, we have turned $l$ and $l_s$ in the unit cm.

For parameters in \ffit, we have
\eqn\fffit{d=1.18^{+1.15}_{-0.71}\times 10^{10},\quad
c=2.70^{+0.15}_{-0.35}\times 10^8.}
Thus for $t_*=10^{-32}$s, $l\sim 10^{-21}$cm, $l_s\sim 10^{-25}$cm. 
$l_s$ is larger than the Planck scale by 8 orders. 
For $t_*=10^{-38}$s, the earliest possibility \all, $l_s\sim 10^{-31}$cm,
larger than the Planck scale only by two orders. For parameters in
\sfit,
\eqn\sffit{d=1.14^{+0.86}_{-0.62}\times 10^{10},\quad
c=2.70^{+0.21}_{-0.31}\times 10^8,}
not much different from those in \fffit.

\bigskip
{\vbox{{\epsfxsize=10cm
        \nobreak
    \centerline{\epsfbox{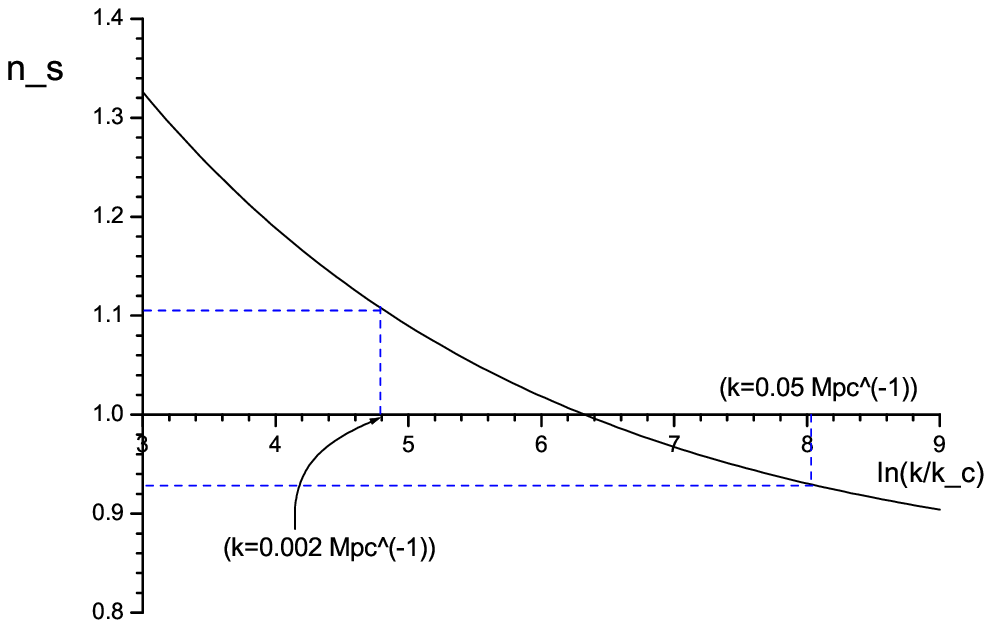}}
        \nobreak\bigskip
    {\raggedright\it \vbox{
{\bf Figure 1.}
{\it $n_s$ as a function of $\ln k$. Parameters are determined using data at
$k=0.05\hbox{Mpc}^{-1}$.}
 }}}}
    \bigskip}

{\vbox{{\epsfxsize=10cm
        \nobreak
    \centerline{\epsfbox{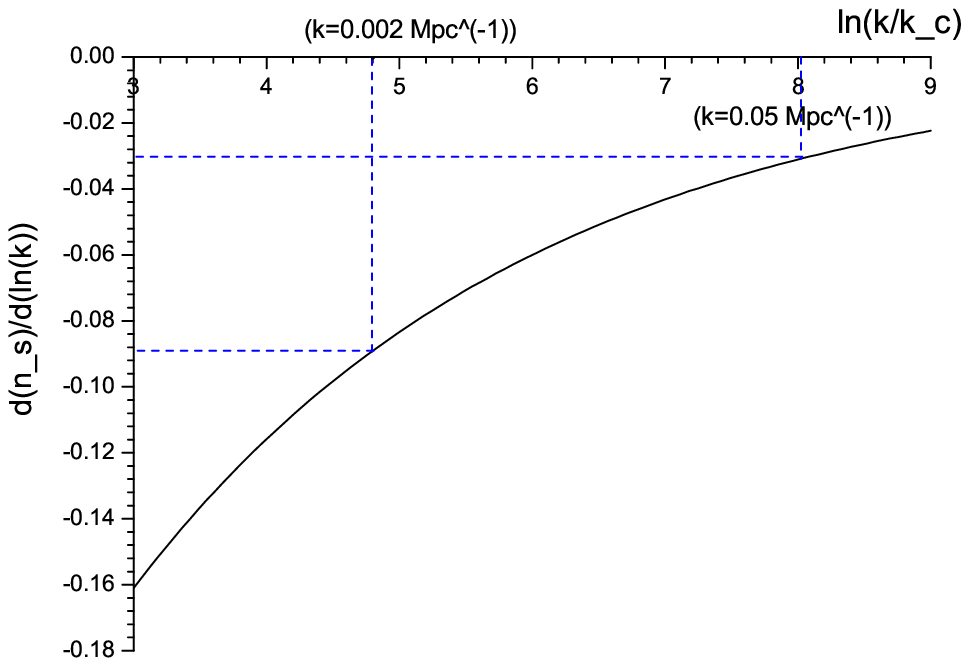}}
        \nobreak\bigskip
    {\raggedright\it \vbox{
{\bf Figure 2.}
{\it ${dn_s\over d\ln k}$ as a function of $\ln k$. Parameters are determined using data at
$k=0.05\hbox{Mpc}^{-1}$.}
 }}}}
    }

To summarize, we have found that the noncommutative inflation model
accommodates the recent WMAP data nicely, and if this model possesses
any truth about string theory, string theory can already be tested
in the observational cosmology, this is very exciting. Although the WMAP
results on the spectral index and its running are obtained by combining
other experiments and are not yet very conclusive, we believe that 
future refined results will offer better opportunity for testing whether
spacetime uncertainty is a viable physical model for inflation.

Acknowledgments. 
This work was supported by a grant of NSC, and by a 
``Hundred People Project'' grant of Academia Sinica and an outstanding
young investigator award of NSF of China. This work was done during a
string workshop organized by ICTS (Interdisciplinary Center of Theoretical 
Sciences), its hospitality is gratefully acknowledged.

\listrefs
\end